\documentclass[sigconf]{acmart}

\usepackage{amsmath}
\usepackage[normalem]{ulem}
\usepackage{hhline}
\usepackage{colortbl}
\usepackage{color}
\usepackage{subfig}
\usepackage{booktabs,makecell, multirow, tabularx}
\usepackage{fancyhdr}
\definecolor{Black}{rgb}{0,0,0}
\AtBeginDocument{%
  }

\setcopyright{none}
\copyrightyear{2023}
\acmYear{2023}
\acmDOI{}

\acmConference[Arxiv]{}{preprint}{2023}
\acmPrice{0.0}
\acmISBN{978-1-4503-9236-5/23/10}
\renewcommand\footnotetextcopyrightpermission[1]{}
\begin{document}

\title{GCNSLIM: Graph Convolutional Network with Sparse Linear Methods for E-government Service Recommendation}


\author{Lingyuan Kong}
\affiliation{%
	\institution{School of Information Management}
	\institution{Nanjing University}
	\city{Nanjing}
	\country{China}}
\email{konglingyuan@smail.nju.edu.cn}
\orcid{}

\author{Hao Ding}
\affiliation{%
  \institution{School of Information Management}
  \institution{Nanjing University}
  \city{Nanjing}
  \country{China}}
\email{dinghao@smail.nju.edu.cn}
\orcid{0000-0003-3528-5686}

\author{Guangwei Hu}
\authornote{Corresponding Authors}
\affiliation{%
	\institution{School of Information Management}
	\institution{Nanjing University}
	\city{Nanjing}
	\country{China}}
\email{hugw@nju.edu.cn}
\orcid{0000－0003－1303－363X}

\renewcommand{\shortauthors}{Lingyuan Kong et al.}

\begin{abstract}
Graph Convolutional Networks (GCNs) have made significant strides in Collaborative Filtering (CF) recommendations. However, existing GCN-based CF methods are mainly based on matrix factorization (MF) and incorporate some optimization techniques (e.g., contrastive learning) to enhance performance, which are not enough to handle the complexities of diverse real-world recommendation scenarios. E-government service recommendation is a crucial area for recommendation research as it involves rigid aspects of people’s lives. However, it has not received adequate attention in comparison to other recommendation scenarios like news and music recommendation. We empirically find that when existing GCN-based CF methods are directly applied to e-government service recommendation, they are limited by the MF framework and showing poor performance. This is because MF’s equal treatment of users and items is not appropriate for scenarios where the number of users and items is unbalanced. In this work, we propose a new model, GCNSLIM, which combines GCN and sparse linear methods (SLIM) instead of combining GCN and MF to accommodate e-government service recommendation. In particular, GCNSLIM explicitly injects high-order collaborative signals obtained from multi-layer light graph convolutions into the item similarity matrix in the SLIM framework, effectively improving the recommendation accuracy. In addition, we propose two optimization measures, removing layer 0 embedding and adding nonlinear activation, to further adapt to the characteristics of e-government service recommendation scenarios. Furthermore, we propose a joint optimization mode to adapt to more diverse recommendation scenarios. We conduct extensive experiments on a real e-government service dataset and a common public dataset and demonstrate the effectiveness of GCNSLIM in recommendation accuracy and operational performance. Our implementation is available at https://github.com/songkk5/GCNSLIM.
\end{abstract}




\maketitle

\section{Introduction}
Compared to popular recommendation research areas like news \cite{sheu2021knowledge,zhang2022metonr,turcotte2015news}, music \cite{xia2022construction,chang2021music}, and movie \cite{liu2023knowledge,roy2022optimal,goswami2022profiling} recommendation, e-government service recommendation has received less attention and witnessed slower development due to the particularity of its working scenarios and the lack of relevant datasets. However, e-government service recommendation is important for several reasons. Firstly, the increasing number of professional services available on e-government platforms has led to information overload, which hinders the effectiveness of e-government services \cite{guo2007intelligent}; Secondly, the demand for government services is ubiquitous, but people often lack the professional knowledge to accurately access relevant government services \cite{liu2020effective}.
\par E-government service recommendation is a typical implicit feedback recommendation \cite{liu2020effective}, where the graph collaborative filtering method \cite{he2020lightgcn,wang2019neural}, which has achieved great success in general recommendation, can be directly applied. The method models user-item interactions as a graph and applies GCN to learn efficient node representations for recommendation \cite{wang2019neural}. It effectively models the high-order connectivity among users and items, and injects collaborative signals into node representations in an explicit way \cite{he2020lightgcn}.
\par Despite the effectiveness, directly applying existing neural graph collaborative filtering methods to e-government service recommendation is not always suitable. As is widely known, the e-government service recommendation scenario is characterized by the large number of users and the limited number of items, particularly in countries with large populations like China and India. For example, in 2021, the number of registered users of the Beijing Municipal People’s Government website exceeded 34.51 million, while the number of government services was only 25,193\footnote{https://www.beijing.gov.cn/gongkai/gkzt/2021szfwzndbb/}; In 2021, the number of registered users of Shanghai Municipal People’s Government website reached 64.44 million, while the number of government services was only 3,458\footnote{https://www.shanghai.gov.cn/2021wzndbb/}. However, the graph collaborative filtering methods based on matrix factorization cannot effectively model this feature due to its equal treatment of the large number of users and the limited number of items, leading to suboptimal recommendation accuracy and long training time.
\par In contrast, sparse linear methods (SLIM)-based models, such as SLIM \cite{ning2011slim}, ADMMSLIM \cite{steck2020admm}, EASE \cite{steck2019embarrassingly}, etc., show convincing performance in the aforementioned scenarios. Unlike MF-based models that simultaneously train the user preference matrix and the item feature matrix, SLIM-based models train only one item similarity matrix. Since the data sparsity problem does not affect the uncertainty in estimating the item similarity matrix when the number of users is large enough \cite{steck2019embarrassingly}, SLIM-based methods can achieve significantly better recommendation accuracy than MF-based methods. Moreover, SLIM-based methods significantly reduce model parameters, which results in faster training speed compared to MF-based methods.
\par In this work, we explore the advantages of combining graph convolutional networks and sparse linear methods, and propose a novel model called GCNSLIM for effective e-government service recommendation. First, we build the infrastructure, taking the product of the item embedding matrix and its transpose matrix as the item similarity matrix, removing the item self-similarity by constraining its diagonals to 0, and then multiplying it with the user-item interaction matrix for prediction, which completes the transformation from MF to SLIM. Second, we enhance the architecture by incorporating multiple light graph convolutional layers on the infrastructure. This enables the propagation of item embeddings on the user-item interaction graph, and the final item embeddings can be obtained through layer connections. The final item embeddings contain high-order collaborative signals, effectively improving the accuracy of the subsequent item similarity matrix calculation. Additionally, we take two optimization measures on the GCN enhanced architecture — removing layer 0 embeddings and adding nonlinear activation. The former further reduces the impact of item self-similarity, and the latter further adapts to the strong centrality of item nodes in the e-government service recommendation scenarios. Furthermore, we also propose a joint optimization mode to flexibly balance the advantages of SLIM and MF to adapt to more diversified recommendation scenarios.
\par The main contributions of our work are summarized as follows:
\par (1) We emphasize the importance of e-government service recommendation and propose our GCNSLIM model to address the particularity of e-government service recommendation scenarios.
\par (2) To the best of our knowledge, our work is the first to combine graph neural networks and sparse linear methods (SLIM) for recommendation systems. It provides a new perspective to extend graph-based collaborative filtering.
\par (3) We conduct extensive experiments on a real e-government service dataset and a common public dataset, and empirically demonstrate that the proposed method outperforms the state-of-the-art baseline methods.

\section{RELATED WORK}
We review existing works on graph-based collaborative filtering, SLIM and its variants, and e-government service recommendations, which are most relevant to this work. Here we emphasize their differences from our GCNSLIM.
\subsection{Graph Collaborative Filtering}
In recent years, graph neural networks have achieved remarkable success in general recommendation tasks \cite{he2020lightgcn,berg2018graph,wu2021self}. This is primarily due to their ability to capture high-order co-signals in user-item interaction histories while constructing embeddings \cite{wang2019neural}. NGCF \cite{wang2019neural} is a typical earlier work that introduces graph neural networks for general recommendation. It first takes ID mapping vectors as initial embeddings of users and items, then propagates embeddings layer by layer based on the user-item bipartite graph, and finally stitches the embeddings of each layer together to obtain the final embeddings for dot-product interaction prediction. LightGCN \cite{he2020lightgcn} simplifies NGCF and achieves better recommendation performance, by removing invalid feature transformation and nonlinear activation in NGCF, and simplifying the implementation of self-connection by improving the layer combination method. HMLET \cite{kong2022linear} further proves that nonlinear activation is not completely invalid, and proposes a mixed linear and nonlinear graph neural network-based method which uses a gating mechanism to dynamically determine whether each node propagates linearly or nonlinearly in a certain layer. To mitigate the negative effects of popularity bias in collaborative filtering, contrastive learning has been introduced in graph collaborative filtering, which aims to construct more uniformly distributed user and item embeddings, and has shown promising results \cite{wu2021self,yu2022graph,lin2022improving}.
\par Despite the great success, we argue that the graph collaborative filtering methods are based on matrix factorization, which is not enough to adapt to diverse application scenarios. The matrix factorization framework decomposes the user-item interaction matrix into a user embedding matrix and an item embedding matrix in the same vector space, and then uses the final embeddings for dot-product interaction prediction. Particularly, equally treating user embeddings and item embeddings makes it challenging for such a framework to fully leverage the capabilities of itself and optimization techniques in scenarios where the number of users and items is unbalanced.
\subsection{SLIM and Variants}
Unlike the matrix factorization model, the SLIM model \cite{ning2011slim} only trains an item similarity matrix, and makes predictions by dot product between the final item similarity matrix and the user-item interaction matrix. It is a simple linear model that combines neighborhood-based and model-based collaborative filtering methods \cite{mao2021simplex}. The objective function it learns is shown in equation \ref{eq1} and \ref{eq2} .
\begin{equation}\label{eq1}
\underset{B}{min}\frac{1}{2}\left\| {X - XB} \right\|_{F}^{2} + \frac{\beta}{2}\left\| B \right\|_{F}^{2} + \lambda\left\| B \right\|_{1}
\end{equation}
\begin{equation}\label{eq2}
	s.t.~~~~~~~~~~~~~B \geq 0,~~{\rm diag}(B) = 0
\end{equation}
where $X$ represents the user-item interaction matrix, and $B$ represents the item similarity matrix. $\left\| \cdot \right\|_{F}$ and $\left\| \cdot \right\|_{1}$ denote the Frobenius-norm and L1-norm, respectively. $\beta$ and $\lambda$ are balancing weights for the regularization terms.
\par In scenarios where the number of users is significantly greater than the number of items, the common data sparsity problem (i.e. the low interaction of a single user) will not affect the accuracy of estimation $B$ \cite{steck2019embarrassingly}, and can significantly improve the training speed. LRec \cite{sedhain2016effectiveness} replaces the squared loss of SLIM with a logistic loss, removes the weight matrix’s two constraints and L1 norm regularization, and learns the user similarity matrix instead of the item similarity matrix. GLSLIM \cite{christakopoulou2016local} trains different SLIM local models for different subsets of users, and then combines the prediction results of local models and global models for recommendation. $\rm EASE^{R}$ \cite{steck2019embarrassingly} removes the non-negative constraint and L1 norm regularization based on SLIM, and derives a closed solution to improve the model training efficiency. ADMMSLIM \cite{steck2020admm} optimizes the original SLIM target by using the alternating direction method of the multiplier ADMM, which allows flexible enabling or disabling of various constraints and regularization items in the original SLIM objective, thus improving the recommendation accuracy.
\par Despite the effectiveness, the aforementioned SLIM-based models do not adequately mine user-item interaction data, without explicitly exploiting the high-order co-signals within them. Moreover, none of the above methods uses deep learning to improve SLIM.
\subsection{E-government Service Recommendation}
Most e-government service recommendation research focuses on enhancing tradtional collaborative filtering or content-based algorithms to better adapt to the e-government recommendation domain. Abdrabbah et al. \cite{abdrabbah2016dynamic} propose a dynamic community-based e-government service recommendation method, which searches for more accurate similar items to optimize recommendations based on the dynamic service community generated from citizen history score and the semantic community of expert predefined classification. Huang et al. \cite{huang2017government} propose an e-government service recommendation algorithm based on community and association sequence mining, which improves the diversity of recommendation results by prefiltering recent user sets, and improves the recommendation accuracy by associative sequence mining of services. Xu et al. \cite{xu2019government} propose an e-government service recommendation method combining probabilistic semantic clustering analysis and improved collaborative filtering, which improves the semantic modeling of users and services. Ayachi et al. \cite{ayachi2016proactive} propose an active and passive e-government service recommendation method, providing passive recommendations by combining citizen needs and contextual information formulated by questionnaires, and providing active recommendations by mining user portrait information from social media. The government service recommendation methods discussed above have improved traditional collaborative filtering and content-based algorithms from various aspects. However, they are still essentially traditional recommendation algorithms which don’t fully utilize machine learning techniques.
\par In recent years, some e-government service recommendation studies have begun to adopt machine learning-based methods, which further promote the development of this field by improving recommendation accuracy and interpretability. Liu et al. \cite{liu2020effective} propose a time-decayed Bayesian personalized ranking e-government service recommendation algorithm, which combines the influence of time series and Bayesian personalized ranking algorithm to improve the accuracy of government service recommendation. Sun et al. \cite{sun2021enhanced} propose a negative items mixed collaborative filtering method to enhance the CF-based e-government service recommendation which mixes the potential features of negative items and positive items to effectively simulate the connectivity on the interaction graph and capture high-level features. Sun et al. \cite{sun2022user} propose a matrix factorization model based on user dynamic topology information for e-government service recommendation which applies topological sequence to the recommendation system for the first time, providing a new perspective to enrich user representation. In general, the number of government service recommendation methods based on deep learning is relatively small, and there is no research on government service recommendation combined with advanced deep learning technologies such as GNN and SAN \cite{wang2020linformer,shaw2018self}.

\section{GCNSLIM}
\begin{figure}[h]
	\centering
	\includegraphics[scale=0.45]{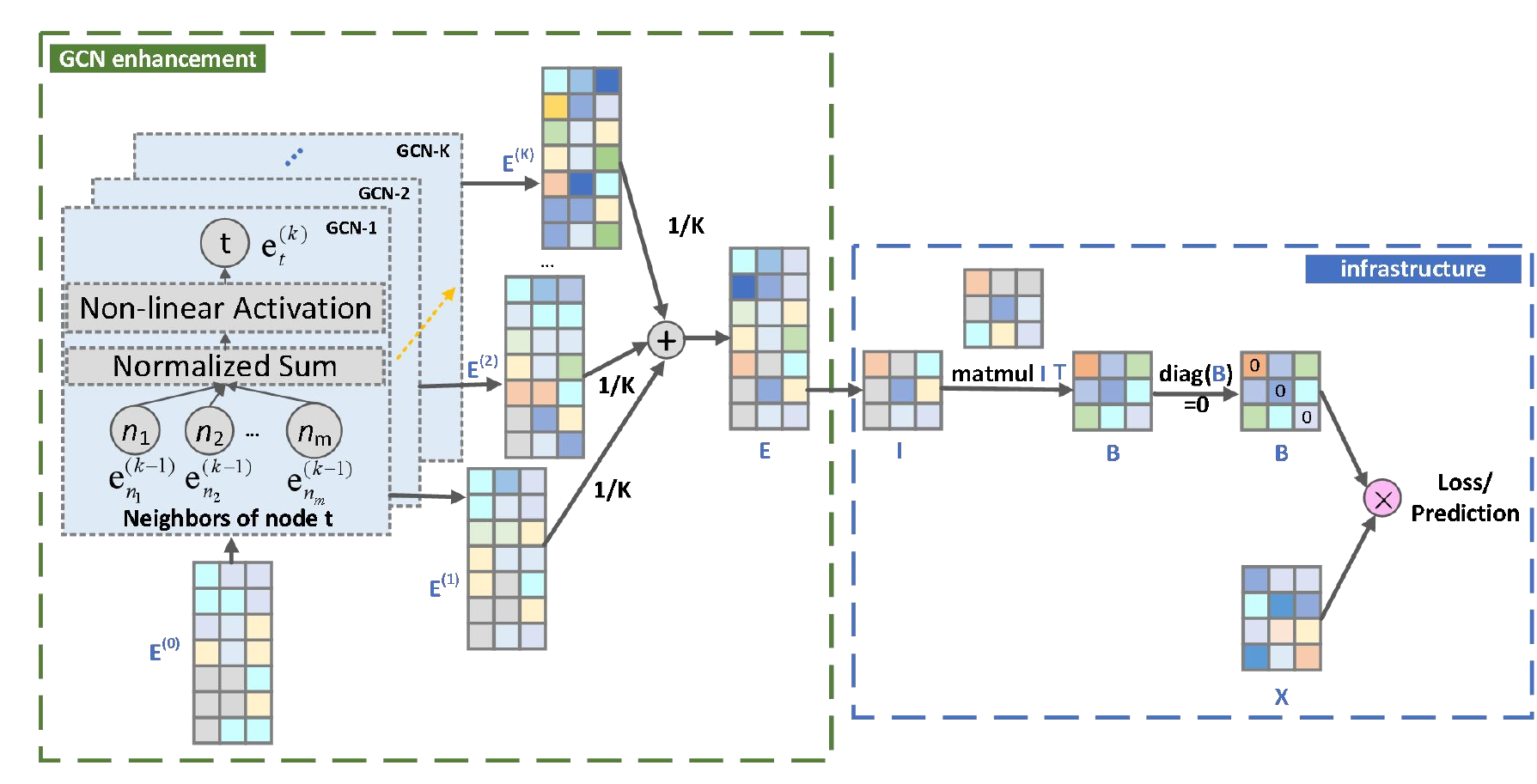}
	\caption{The Model Architecture of GCNSLIM.}
	\Description{The Model Architecture of GCNSLIM.}
	\label{fig1}
\end{figure}

We now present our GCNSLIM model, as shown in Figure \ref{fig1}. The introduction to the architecture is divided into four parts: infrastructure, GCN enhancement, joint optimization, and model discussion.
\subsection{Infrastructure}
The existing graph collaborative filtering models are all evolved under the matrix factorization framework, and GCN is mainly responsible for optimizing embeddings. But under the SLIM framework, there is no concept of embedding but only the concept of item similarity matrix. A simple solution is to take the product of the item embedding matrix and its own transpose matrix as the item similarity matrix. This simple operation naturally converts the similarity between item embeddings under the MF framework to specific similarity values in the item similarity matrix under the SLIM framework, and explicitly guarantees the symmetry of the item similarity matrix. Furthermore, according to the zero diagonal constraint of SLIM, the diagonal of the converted item similarity matrix is zeroed. Its purpose is to avoid the trivial solution, i.e., $B$ equals the identity matrix. Note that we do not employ non-negativity constraints and L1-norm regularization here. This is because previous literatures \cite{steck2020admm,steck2019embarrassingly} have found that in scenarios with a large number of users, the non-negative constraint of SLIM tends to reduce the ranking accuracy, and L1-norm regularization has limited effect on the ranking accuracy while reducing the computational efficiency.
\par Finally, we learn item embeddings using the convex objective such as Equation \ref{eq3}:
\begin{equation}\label{eq3}
\underset{I~~~~~~~~~~}{min~~~~~~~~}\left\| {X - X\left( II^{{\rm T}} - {\rm diag}(II^{{\rm T}}) \right)} \right\|_{F}^{2} + \lambda\left\| I \right\|_{F}^{2}
\end{equation}
where $I$ and $I^{T}$ represent the overall item embedding matrix and its transpose matrix, respectively, and the rest of the symbols have the same meanings as equation \ref{eq1} and \ref{eq2}. Following SLIM, we learn weights by minimizing the squared loss. Literatures \cite{steck2019embarrassingly,liang2018variational} point out that multinomial likelihood leads to better ranking accuracy than logistic likelihood (logarithmic loss) or Gaussian likelihood (squared loss). We leave the exploration as future work.
\par SLIM-like methods usually decompose the optimization problem into independent parallel problems \cite{ning2011slim,sedhain2016effectiveness,levy2013efficient}. The calculation of each parallel problem uses once elastic network regression. The column of the user-item interaction matrix $X$ corresponding to the current item is used as the target value, and the $X$ after the column is set to 0 is used as the training data. Through multiple iterations of training, the similarity between the current item and other items can be obtained. Such processing is to treat all 0 points in $X$ as negative samples and fully sample them, which ignores 0 points representing unobserved positive sample points. To this end, we replace 0-point full negative sampling with 0-point random negative sampling (the negative sampling rate is optimized on the validation set as a hyperparameter).

\subsection{GCN Enhancement}
The GCNSLIM infrastructure has limitations in mining user-item interaction data and using structural information explicitly. Introducing GCN enhancement to the GCNSLIM infrastructure can explicitly inject high-order connectivity in user-item interaction data into item embeddings, making structurally related item embeddings more similar, thereby obtaining a more accurate item similarity matrix to improve recommendation performance. Specifically, we define the graph convolution operation in GCNSLIM as Equations \ref{eq4} and \ref{eq5}:
\begin{equation}\label{eq4}
e_{u}^{(k + 1)} = LeakyReLU\left( {\sum\limits_{i \in N_{u}}{\left. \frac{1}{\sqrt{\left| N_{u} \right|}\sqrt{\left| N_{i} \right|}}e_{i}^{(k)} \right)}} \right.
\end{equation}
\begin{equation}\label{eq5}
e_{i}^{(k + 1)} = LeakyReLU\left( {\sum\limits_{u \in N_{i}}{\left. \frac{1}{\sqrt{\left| N_{i} \right|}\sqrt{\left| N_{u} \right|}}e_{u}^{(k)} \right)}} \right.
\end{equation}
where $e_u^{(k+1)}$ and $e_i^{(k+1)}$ denote the representations of user $u$ and item $i$ obtained after going through the $k+1^{\rm th}$ embedding propagation layer, respectively. $N_{u}$ represents the interactive item set of user $u$, and $N_{i}$ represents the interactive user set of item $i$. This graph convolution operation is adding LeakyReLU nonlinear activation to the graph convolution operation of LightGCN \cite{he2020lightgcn}. Although LightGCN claims that nonlinear activation has no positive impact on collaborative filtering, HMLET \cite{kong2022linear} proves that nonlinear activation is not completely ineffective, and finds that nodes with strong centrality are more suitable for nonlinear propagation. In the e-government service scenario, the number of items is usually much smaller than the number of users, and most item nodes are important nodes with strong centrality. Since the main optimization parameters of GCNSLIM are item embeddings, adding an appropriate nonlinear activation function can significantly improve the recommendation performance.
\par After $K(K>0)$ layer propagation, we use the weighted sum function as the output function to combine the output embeddings of all graph convolutional layers and obtain the final representations as Equations (6) and (7):
\begin{equation}\label{eq6}
e_{u} = \frac{1}{K}{\sum\limits_{k = 1}^{K}e_{u}^{(k)}}
\end{equation}
\begin{equation}\label{eq7}
e_{i} = \frac{1}{K}{\sum\limits_{k = 1}^{K}e_{i}^{(k)}}
\end{equation}
where $e_{u}$ and $e_{i}$ denote the final embeddings of user $u$ and item $i$. The layer combination here is to remove the layer 0 embedding from the layer combination in LightGCN. Combining layer 0 embeddings in layer connections means considering the item self-similarity, but removing item self-similarity in SLIM-like models is critical \cite{steck2019embarrassingly}.
\par To provide an overall view of embedding propagation and transformation, and to facilitate batch implementation, we present the matrix form of embedding propagation and transformation in GCNSLIM. Equations \ref{eq8} and \ref{eq9} show the rules for layer-by-layer propagation of embeddings in GCN enhancement, removing the feature transformation matrix compared to NGCF \cite{wang2019neural}, and adding a nonlinear activation function compared to LightGCN \cite{he2020lightgcn}.
\begin{equation}\label{eq8}
E^{(k + 1)} = LeakyReLU\left( \left( {D^{- \frac{1}{2}}AD^{- \frac{1}{2}}} \right)E^{(k)} \right),
\end{equation}
\begin{equation}\label{eq9}
A = \begin{bmatrix}
	0 & X \\
	X^{{\rm T}} & 0 \\
\end{bmatrix}
\end{equation}
where $X$ is the user-item interaction matrix, $A$ is the adjacency matrix, $D$ is the degree matrix of $A$, and $E^{(k + 1)}$ denotes the overall embedding matrix (including overall user embeddings and item embeddings) output by layer $k+1$. Finally, the final embedding matrix output by the GCN enhancement is shown in Equations \ref{eq10}, \ref{eq11} and \ref{eq12}:
\begin{equation}\label{eq10}
E = \frac{1}{K}\left( E^{(1)} + E^{(2)} + \ldots + E^{(K)} \right),
\end{equation}
\begin{equation}\label{eq11}
E = \frac{1}{K}\left( {L\left( {\overset{\sim}{A}E}^{(0)} \right) + L\left( {\overset{\sim}{A}E}^{(1)} \right) + \ldots + L\left( {\overset{\sim}{A}E}^{({K - 1})} \right)} \right),
\end{equation}
\begin{equation}\label{eq12}
E = \frac{1}{K}\left( {F\left( E^{(0)} \right) + F^{2}\left( E^{(0)} \right) + \ldots + F^{K}\left( E^{(0)} \right)} \right),
\end{equation}
where $L$ represents the LeakyReLU activation function, $\widetilde{A}=D^{-\frac{1}{2}}AD^{-\frac{1}{2}}$ is a symmetric normalized matrix, and $F(E^{(0)}) = L(\widetilde{A}E^{(0)})$.
\par Further, as shown in Equation \ref{eq13}, the last $N$ ($N$ is the total number of items) embeddings are taken out from the final embedding matrix $E$ to obtain the overall item embedding matrix $I$. As shown in Equations \ref{eq14} and \ref{eq15}, we multiply $I$ by its own transposed matrix $I^{{\rm T}}$ to obtain the item similarity matrix $B$, and set the diagonal of $B$ to 0 to remove the item self-similarity.
\begin{equation}\label{eq13}
I = E\lbrack-N:\rbrack
\end{equation}
\begin{equation}\label{eq14}
B = II^{{\rm T}}
\end{equation}
\begin{equation}\label{eq15}
{\rm diag}(B) = 0.
\end{equation}
\par Finally, the model prediction is defined as the product of the user-item interaction matrix $X$ and the item similarity matrix $B$ as shown in Equation \ref{eq16}:
\begin{equation}\label{eq16}
\hat{X} = XB
\end{equation}
which is used to generate recommendation scores.
\subsection{Joint Optimization}
\begin{figure}[h]
	\centering
	\includegraphics[scale=0.45]{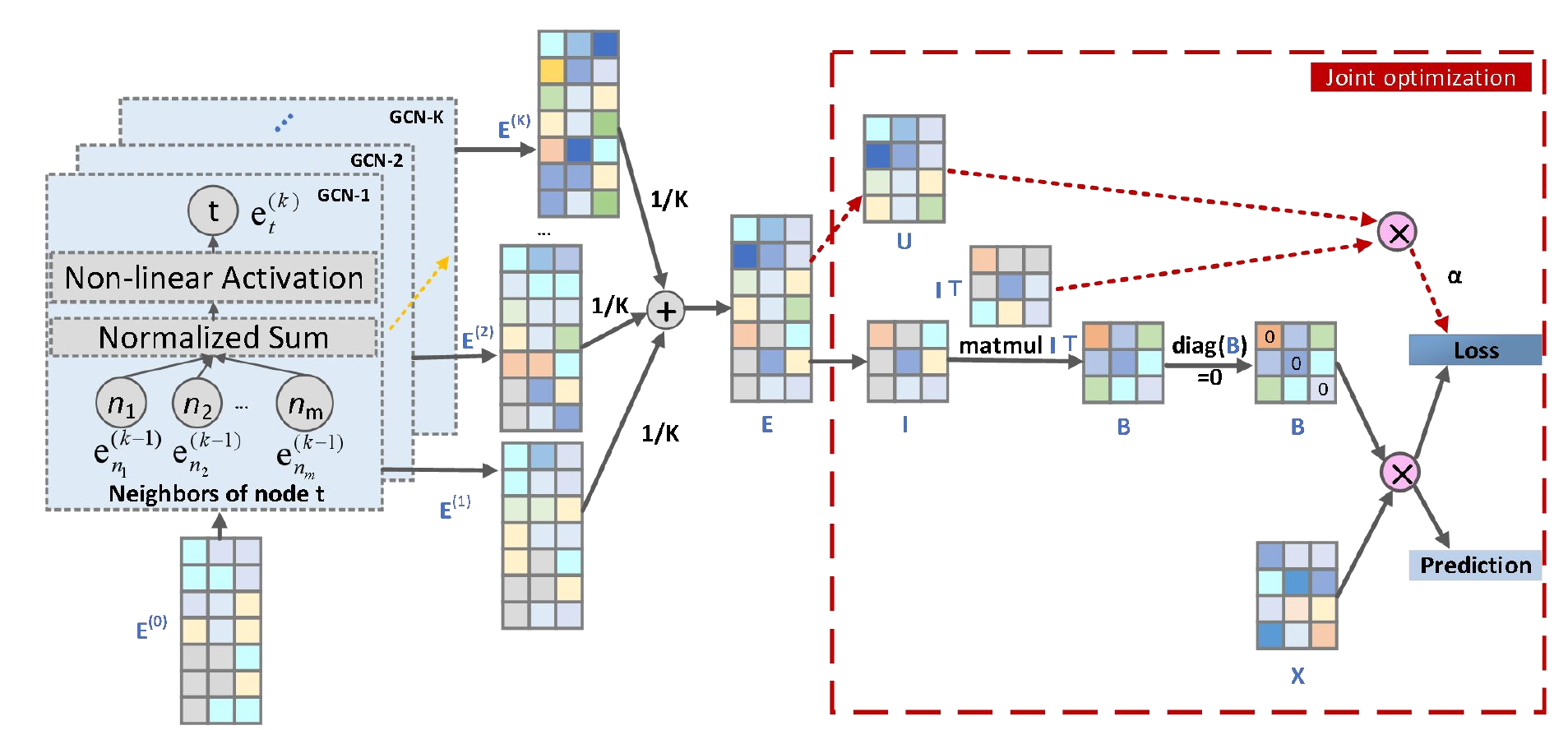}
	\caption{The Joint Optimization Mode of GCNSLIM.}
	\Description{The Joint Optimization Mode of GCNSLIM.}
	\label{fig2}
\end{figure}

To make the proposed GCNSLIM not only serve the scenarios where the user/item sample size distribution is skewed such as e-government service recommendation, but also flexibly adapt to other scenarios where the user/item sample size distribution is more balanced, we introduce joint optimization on the above GCNSLIM architecture, as shown in Figure \ref{fig2}. Specifically, based on the GCNSLIM version in Figure \ref{fig1}, a new supervised learning task is added to multiply the final user embedding matrix and the transpose matrix of the final item embedding matrix to assist in the calculation of the joint loss. Consequently, GCNSLIM is optimized for two supervised learning tasks at the same time, balancing the advantages of both SLIM and MF. By properly adjusting the joint optimization weight $\alpha$ on the validation set, it can flexibly adapt to multiple application scenarios. Specifically, when $\alpha$ is smaller, GCNSLIM is more suitable for recommendation scenarios with a large number of users and a small number of items; when $\alpha$ is larger, GCNSLIM is more suitable for recommendation scenarios with balanced user/item volume.
\par The total loss of GCNSLIM after adding the joint optimization structure can be expressed as Equation \ref{eq17}:
\begin{equation}\label{eq17}
L = \left\| {X - X\left( II^{{\rm T}} - {\rm diag}\left( {II^{{\rm T}}} \right) \right)} \right\|_{F}^{2} + \alpha\left\| {X - UI^{{\rm T}}} \right\|_{F}^{2} + \lambda\left\| E^{(0)} \right\|_{F}^{2}
\end{equation}
where $U$ represents the overall user embedding matrix, and the rest of the symbols are the same as above.
\subsection{Model Discussion}
\subsubsection{Relation with IGCCF}
We note a recent paper \cite{d2023item} that proposes IGCCF that, like our GCNSLIM, emphasizes utilizing graph convolutions to inject high-order collaborative signals into item embeddings. It builds an item-item graph based on user-item interactions, and propagates item embeddings on it, then builds user embeddings as the weighted sum of associated item embeddings, and finally performs a dot product using user embeddings and item embeddings for prediction. Unlike IGCCF which explicitly learns high-order collaborative signals to item embeddings on the item-item graph, our GCNSLIM still optimizes item embeddings on the user-item bipartite graph and leverages the SLIM framework to drive the model to focus more on learning item embeddings (compared to user embeddings). Since GCNSLIM uses user interaction records and item similarity matrix for prediction, it has the same ability to process dynamic graphs as IGCCF and is more efficient. Specifically, when existing users generate new interactions or new users enter the system and start to generate interactions, GCNSLIM can directly make new predictions based on the new interactions and the previously trained item similarity matrix, without the need for retraining like LightGCN, or building user embeddings before predictions like IGCCF. In addition, although GCNSLIM does not reduce the space complexity of the baseline model LightGCN like IGCCF, GCNSLIM is simpler and more efficient and retains the possibility of graph collaborative optimization combined with user context information.
\subsubsection{User-side Expansion}
\begin{figure}[h]
	\centering
	\includegraphics[scale=0.45]{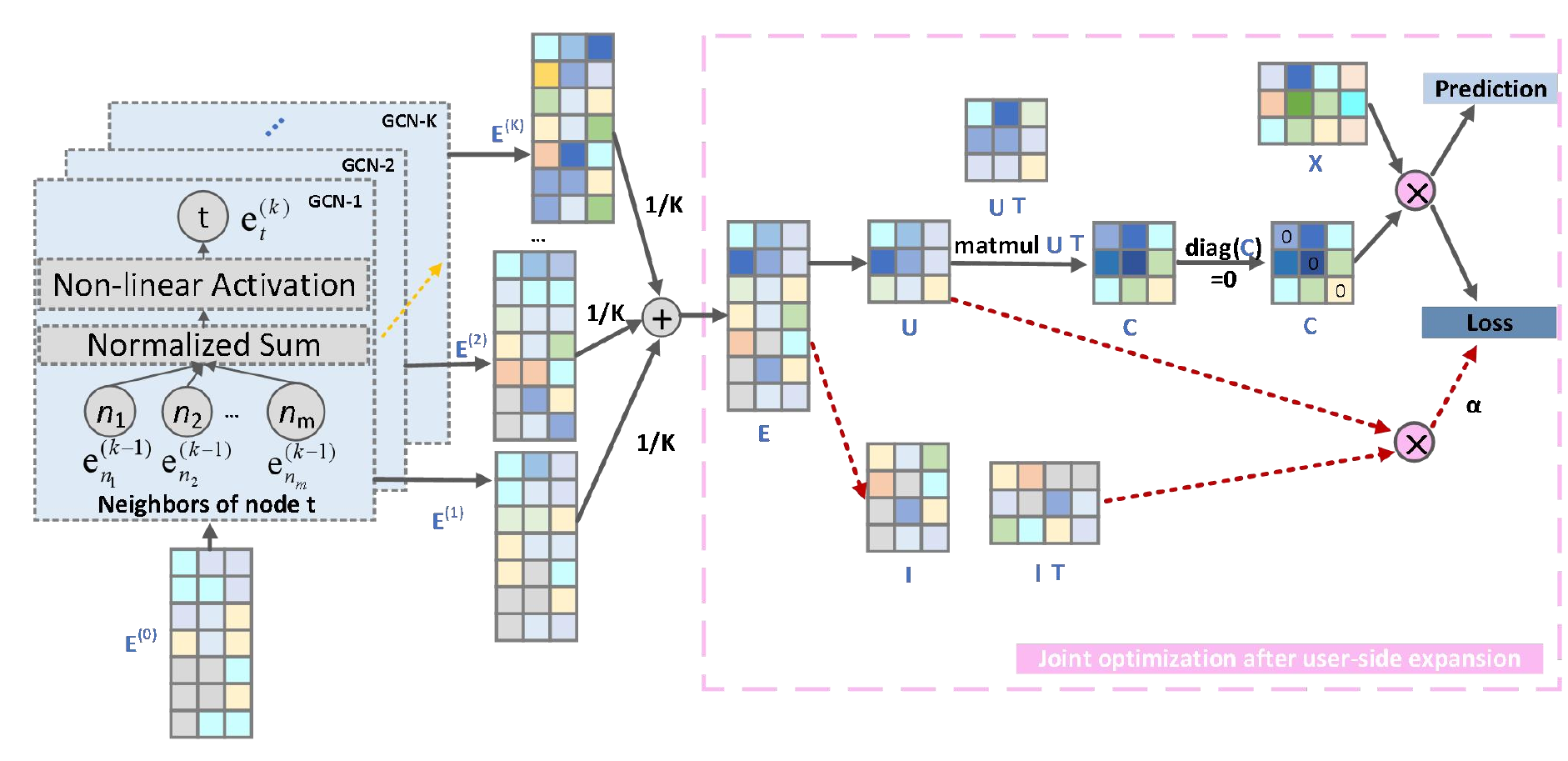}
	\caption{The User-side Expansion Version of GCNSLIM.}
	\Description{The User-side Expansion Version of GCNSLIM.}
	\label{fig3}
\end{figure}

GCNSLIM has strong scalability and can be extended to the user side. For some special application scenarios with small user groups (such as niche music circles), the user-item interaction data may have the characteristics of large item volume and small user volume (contrary to the e-government service recommendation scenarios), which cannot be directly and properly handled by existing MF-like methods or SLIM-like methods. However, inspired by user-based collaborative filtering methods \cite{jain2020emucf,koohi2016user} GCNSLIM can adapt to the recommendation requirements of the scenarios by expanding user-side interaction modeling. The specific process is to symmetrically exchange the user roles and item roles of the jointly optimized version of GCNSLIM in Figure \ref{fig2}. As shown in Figure \ref{fig3}, on one side, the overall user embedding matrix obtained after GCN propagation is multiplied by its own transpose matrix to get the user similarity matrix, and the diagonal of the user similarity matrix is then set to 0 before multiplying it by the user-item interaction matrix for prediction and joint optimization loss; On the other side, the overall user embedding matrix and the transpose matrix of the overall item embedding matrix are multiplied for the joint optimization loss.
\par Similarly, the total loss of the user-side expansion version of GCNSLIM can be expressed as Equation \ref{eq18}:
\begin{equation}\label{eq18}
L = \left\| {X - \left( UU^{{\rm T}} - {\rm diag}\left( {UU^{{\rm T}}} \right) \right)X} \right\|_{F}^{2} + \alpha\left\| {X - UI^{{\rm T}}} \right\|_{F}^{2} + \lambda\left\| E^{(0)} \right\|_{F}^{2}
\end{equation}
\par The user-side expansion version of GCNSLIM also has the ability to process dynamic graphs. Specifically, when existing items generate new interactions or new items enter the system and start to generate interactions, GCNSLIM can directly make new predictions based on the new interactions and the previously trained user similarity matrix. The user-side expansion of GCNSLIM is theoretically feasible, but since this paper focuses on e-government service recommendation, the relevant empirical research will be left to future work.

\section{EXPERIMENTS}
To verify the superiority of GCNSLIM and reveal the reasons for its effectiveness, we conduct extensive experiments on a real-world e-government service dataset and a commonly used public dataset Movielens-1M and answer the following questions:
\begin{itemize}
	\item RQ1: How does GCNSLIM perform as compared with state-of-the-art CF methods?
	\item RQ2: How do different components of GCNSLIM affect GCNSLIM?
	\item RQ3: How do different hyper-parameter settings affect GCNSLIM?
\end{itemize}
\subsection{Dataset Description}
The e-government service dataset Wuhou comes from the open data upon application of the Wuhou District Administrative Approval Bureau in Chengdu\footnote{http://gk.chengdu.gov.cn/openApply/}, which is consistent with the experimental data sources in literatures \cite{liu2020effective,sun2021enhanced,sun2022user}. The source data we received includes more than 4 million records of more than 60,000 citizens in the region on more than 500 government services from October 2015 to August 2019. The data items included in the dataset are user ID, gender, age, item ID, and timestamp. We clean and deduplicate the source data, and use the 10-core setting \cite{he2016vbpr} on this basis to ensure that each user and item has at least 10 interactions. For Movielens-1M, we consider interactions with ratings no less than 3 as positive interactions \cite{d2023item} and adopt the 10-core setting. We present the statistics of each dataset in Table \ref{tab:database}.
\begin{table}[h]
	\caption{Statistics of the datasets}
	\label{tab:database}
	\begin{tabular}{ccccc}
		\toprule
		\textbf{Dataset} & \textbf{\#Users} & \textbf{\#Items} & \textbf{\#Interactions} & \textbf{Density}\\
		\midrule
		Wuhou & 16216 & 437 & 540557 & 0.07628\\
		Ml-1M & 6033 & 3123 & 834449 & 0.04429\\
		\bottomrule
	\end{tabular}
\end{table}

\subsection{Experimental Settings}
\subsubsection{Evaluation Metrics}
To evaluate the impact of top-N recommendation and preference ranking, we adopt two widely used evaluation metrics Recall@N and NDCG@N. Considering that the two datasets are relatively small and the convenience requirements of the e-government service scenarios make them focus on the top of the recommendation list, N is set to 10. Following previous work \cite{he2020lightgcn,wang2019neural}, we adopt the full ranking strategy, which is to sort all candidate items that the user has not interacted with.
\subsubsection{Baselines}
To demonstrate the effectiveness, we compare our GCNSLIM with the following models.
\begin{itemize}
	\item \textbf{NGCF}\cite{wang2019neural}: This is a typical graph collaborative filtering model, which greatly follows the standard GCN. We tune the regularization coefficient $\lambda_{2}$ and the number of GCN layers within the suggested range.
	\item \textbf{LightGCN}\cite{he2020lightgcn}: This model is simplified on NGCF, and a light graph convolution is designed for training efficiency and generation ability. Similarly, we adjust the regularization coefficients $\lambda_{2}$ and the number of GCN layers within the suggested range.
	\item \textbf{SLIM}\cite{ning2011slim}: This is the original sparse linear method. We adjust the regularization coefficients $\lambda_{1}$, $\lambda_{2}$ and whether to consider non-negative constraints within the recommended range.
	\item $\bf EASE^{R}$\cite{steck2019embarrassingly}: It removes the non-negative constraint and L1 norm regularization based on SLIM, and derives a closed solution to improve the model training efficiency. We adjust the regularization coefficien $\lambda_{2}$ within the range suggested in the original paper.
	\item \textbf{ADMMSLIM}\cite{steck2020admm}: It optimizes the original SLIM target using the alternating direction method of the multiplier ADMM, allowing flexibility to enable or disable various constraints and regularization terms in the original SLIM target. We adjust the regularization coefficients $\lambda_{1}$, $\lambda_{2}$, and the squared difference penalty term $\rho$ within the range suggested in the original paper.
\end{itemize}
\subsubsection{Parameter Settings}
We implement the proposed model and all baselines using RecBole \cite{zhao2021recbole}, a unified open source framework for developing and replicating recommendation algorithms. To ensure fair comparisons, we optimize all methods using the Adam optimizer and initialize all parameters using the default Xavier distribution. We uniformly set the batchsize to 4096 and the embedding size to 128. We adopt early stopping with the patience of 50 epoch to prevent overfitting, and NDCG@10 is set as the indicator. We set the ratio of training, validation, and testing set to 6:2:2, set the learning rate to 0.001, and fine-tune the number of negative samples corresponding to a positive case at \{1, 2, 3, 4\}. For GCNSLIM, we adjust the joint optimization weight $\alpha$ at \{0.0, 0.01, 0.05, 0.1, 0.5\}, the L2 regularization coefficient $\lambda$ at \{0.1, 0.5, 1.0, 1.5, 2.0, 5.0\}, and the number of GCN layers at \{1, 2, 3\}.

\subsection{Overall Experiments}
Table \ref{tab:OPC} shows the overall performance comparison results. We have the following observations:\\
(1) GCNSLIM consistently outperforms all baselines on both datasets. This demonstrates the rationality and effectiveness of combining graph convolutional networks with sparse linear methods.\\
\begin{table}[htbp]\footnotesize
	\renewcommand\arraystretch{1.3}
	\centering
	\caption{Overall performance comparison}
	\label{tab:OPC}
	\begin{tabular}{c|cc|cc}
		\bottomrule[1.1pt]
		\multirow{2}[1]{*}{Model} & \multicolumn{2}{c|}{Wuhou} & \multicolumn{2}{c}{Ml-1M} \\
		\cline{2-5}    \multicolumn{1}{c|}{} & \multicolumn{1}{p{4.19em}}{Recall@10} & \multicolumn{1}{p{4.19em}|}{NDCG@10} & \multicolumn{1}{p{4.19em}}{Recall@10} & \multicolumn{1}{p{5.19em}}{NDCG@10} \\
		\hline
		SLIM  & 0.5897 & 0.5860 & 0.1706 & 0.3682 \\
		$\rm EASE^{R}$ & 0.5898 & 0.5861 & 0.1714 & 0.3714 \\
		ADMMSLIM & 0.5927 & \underline{0.5875} & 0.1736 & \underline{0.3761} \\
		\hline
		NGCF  & 0.5813 & 0.5783 & 0.1633 & 0.3557 \\
		LightGCN & 0.5879 & 0.5836 & 0.1733 & 0.3743 \\
		\hline
		GCNSLIM($\alpha$=0) & \textbf{0.594} & \textbf{0.5898} & \underline{0.1778} & 0.3752 \\
		GCNSLIM($\alpha$=0.05) & \underline{0.5937} & \textbf{0.5898} & \textbf{0.1786} & \textbf{0.3769} \\
		\toprule[1.1pt]
	\end{tabular}%
	\label{tab:addlabel}%
\end{table}%
(2) On the e-government service dataset, SLIM-like methods are consistently better than MF-like methods. This proves that in the e-government service recommendation scenario, SLIM-like methods have better applicability for modeling the recommendation task with a large number of users and a small number of items.\\
(3) On the Ml-1M dataset, the advanced SLIM-like algorithms GCNSLIM and ADMMSLIM are consistently superior to the advanced MF-like algorithms LightGCN and NGCF, but the advantage distance is shorter than that on the Wuhou dataset. This makes sense, because the number of Ml-1M users is still significantly larger than the number of items, but its data volume distribution is more balanced compared to Wuhou.\\
(4) The performance of GCNSLIM ($\alpha$=0) is better than GCNSLIM ($\alpha$=0.05) on Wuhou and weaker than GCNSLIM ($\alpha$=0.05) on Ml-1M. This shows that GCN combined with pure SLIM is sufficient to cope with e-government service recommendation scenarios with skewed data volume distribution, while joint optimization measures play a greater role in scenarios where data volume distribution is more balanced.\\
(5) On the two datasets, GCNSLIM and LightGCN achieve better performance than NGCF, which reflects that the light graph convolution is still better than the standard graph convolution on the dataset with uneven distribution of the number of users and items. In particular, the poor performance of NGCF on Ml-1M implies that the feature transformation operator will lead to more severe overfitting in scenarios with relatively rich items and interactions.\\
(6) On both datasets, ADMMSLIM and $\rm EASE^{R}$ have improved performance compared to SLIM, and ADMMSLIM’s performance is better than $\rm EASE^{R}$. This indicates that they are effective for SLIM performance optimization, and the flexible automatic control of SLIM target items is significantly superior to the heuristic control.\\
\begin{figure}[h]
	\centering
	\subfloat[\label{fig:4a}]{
		\includegraphics[scale=0.355]{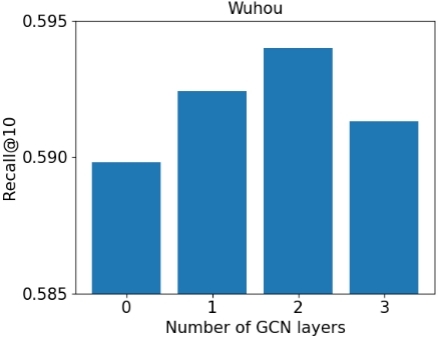}}
	\subfloat[\label{fig:4b}]{
		\includegraphics[scale=0.355]{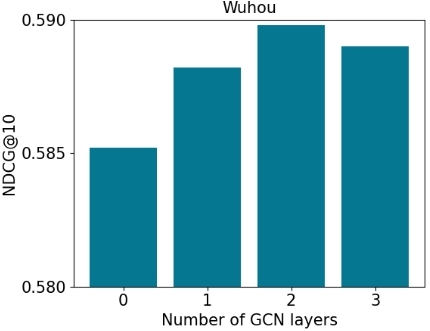}}
	\\
	\subfloat[\label{fig:4c}]{
		\includegraphics[scale=0.34]{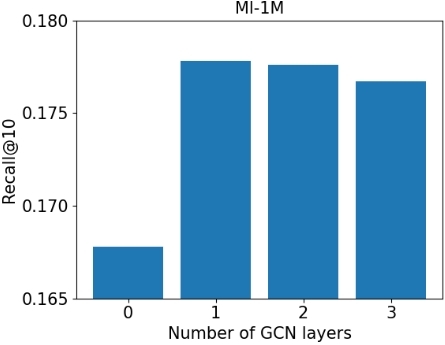}}
	\subfloat[\label{fig:4d}]{
		\includegraphics[scale=0.34]{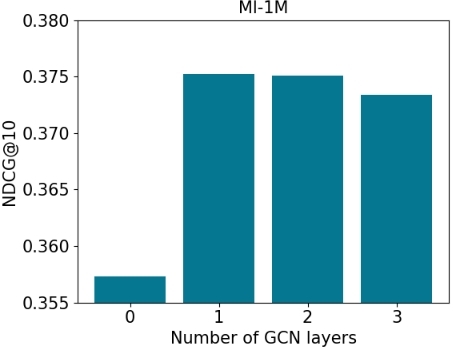}}
	\caption{Effect of graph convolution layers.}
	\label{fig4} 
\end{figure}

\subsection{Ablation Experiments}
\subsubsection{The effectiveness of GCNs}
To investigate whether GCNSLIM can benefit from the high-order connectivity propagated by GCNs, we transform the model depth. In particular, taking $\alpha$=0 as an example, search for the number of layers $K$ in \{0, 1, 2, 3\}, where the $\rm 0^{th}$ layer embeddings are retained when $K$=0. Figure \ref{fig4} shows the results of the experiment, and we have the following observations:\\
(1) In all cases, using GCN components ($K$ > 0) greatly increases the recommendation performance (for example, Recall and NDCG of 1-layer GCN improve by 5.96\% and 5.01\% compared to 0-layer GCN on ML-1M), empirically showing that GCNSLIM can benefit from the collaborative signals propagated by GCNs.\\
(2) With the increasing number of GCN layers, the recommendation performance of GCNSLIM first increases and then decreases in all cases, reaching the best performance when the number of layers is equal to 2 in the Wuhou dataset. This reflects that GCNSLIM considering second-order neighbors can optimally balance mining high-order connectivity and reducing overfitting in the e-government service recommendation dataset, while applying deeper structures may introduce noise into representation learning. The best results are achieved when the number of layers is equal to 1 in the Ml-1M dataset, which implies that the collaborative signals carried by first-order neighbors are sufficient for GCNSLIM to achieve the best performance when the amount of items and interactions are relatively abundant.\\
(3) Comparing Figure \ref{fig4} and Table \ref{tab:OPC}, when changing the number of propagation layers, GCNSLIM is consistently better than non-SLIM-like methods (i.e. NGCF, LightGCN) on the Wuhou dataset. It once again proves the effectiveness of the SLIM framework for e-government service recommendations.
\subsubsection{The effectiveness of SLIM}
\begin{figure}[h]
	\centering
	\subfloat[\label{fig:5a}]{
		\includegraphics[scale=0.28]{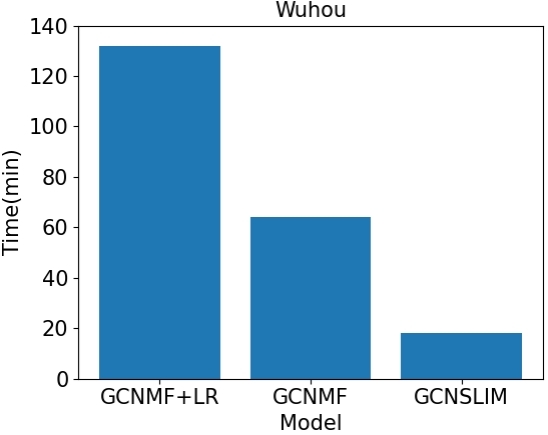}}
	\subfloat[\label{fig:5b}]{
		\includegraphics[scale=0.28]{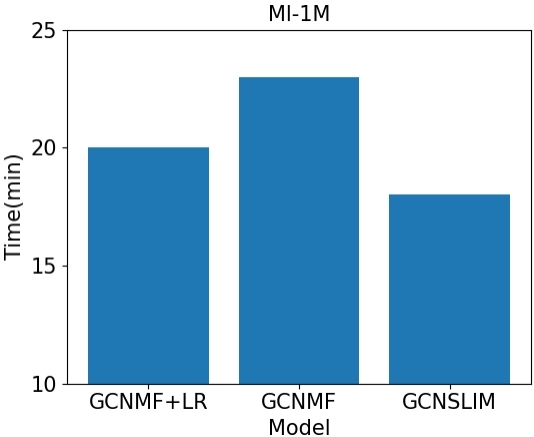}}
	\caption{Training time comparison of GCNSLIM and GCNMFs.}
	\label{fig5} 
\end{figure}

To investigate whether GCNSLIM could benefit from the SLIM framework, we consider a variant of GCNSLIM based on the MF framework, terminated as GCNMF. In particular, for GCNSLIM with 1-layer or 2-layer GCN, we keep the layer 0 embeddings, remove the transpose multiplication and subsequent components, and use the user embeddings and item embeddings for dot product interaction prediction. Moreover, we also consider the GCNMF variant with nonlinear activation, $+$LR means adding nonlinear activation. We show the performance comparison results with GCNSLIM ($\alpha$=0) in Table \ref{tab:PCG}, and show the training time comparison results in Figure \ref{fig5} (the experimental environment is RTX 3090 24G), and we have the following observations:\\
(1) In all cases, GCNSLIM consistently outperforms MF-like variants, empirically showing that GCNSLIM can benefit from the SLIM framework. This shows that the SLIM framework can more accurately model the e-government service recommendation scenarios.\\
(2) Comparing the training time of GCNSLIM, GCNMF and GCNMF+LR to achieve optimal performance on the Wuhou dataset, it can be found that the convergence speed of GCNSLIM is significantly faster than the other two (the training time is reduced by an average of 79.12\%). This shows that the SLIM framework can model e-government service recommendation scenarios more quickly and efficiently.\\
(3) Comparing the performance of GCNMF and GCNMF+LR, it can be observed that the two variants have comparable performance. This implies that the non-linear activation has no significant effect on the MF-like graph collaborative filtering algorithms, which is consistent with the results in the paper \cite{he2020lightgcn}.
\begin{table}[htbp]\footnotesize
    \renewcommand\arraystretch{1.3}
	\centering
	\caption{Performance comparison of GCNSLIM and GCNMFs.}
	\label{tab:PCG}
	\begin{tabular}{c|c|cc|cc}
		\bottomrule[1.1pt]
		\multicolumn{1}{c|}{\multirow{2}[1]{*}{Layer}} & \multirow{2}[1]{*}{Model} & \multicolumn{2}{c|}{Wuhou} & \multicolumn{2}{c}{Ml-1M} \\
		\cline{3-6}          & \multicolumn{1}{c|}{} & \multicolumn{1}{p{4.19em}}{Recall@10} & \multicolumn{1}{p{4.19em}|}{NDCG@10} & \multicolumn{1}{p{4.19em}}{Recall@10} & \multicolumn{1}{p{4.19em}}{NDCG@10} \\
		\hline
		& GCNMF+LR & 0.5894 & 0.5870 & 0.1730 & 0.3716 \\
		\multicolumn{1}{c|}{\#1} & GCNMF & 0.5866 & 0.5840 & 0.1738 & 0.3726 \\
		& GCNSLIM & \textbf{0.5924} & \textbf{0.5882} & \textbf{0.1778} & \textbf{0.3752} \\
		\hline
		& GCNMF+LR & 0.5892 & 0.5879 & 0.1749 & 0.3734 \\
		\multicolumn{1}{c|}{\#2} & GCNMF & 0.5900  & 0.5888 & 0.1748 & 0.3727 \\
		& GCNSLIM & \textbf{0.5940} & \textbf{0.5898} & \textbf{0.1776} & \textbf{0.3751} \\
		\toprule[1.1pt]
	\end{tabular}%
	\label{tab:addlabel}%
\end{table}%

\subsubsection{The effectiveness of optimization measures}
To investigate whether GCNSLIM can benefit from the two proposed optimizations — removing layer 0 embeddings and adding nonlinear activations — we consider GCNSLIM’s variants without the optimizations. Among them, GCNSLIM+0 represents the variant that retains the layer 0 embeddings, GCNSLIM-LR represents the variant that removes LeakyReLU, and GCNSLIM+0-LR represents the variant without both optimizations. In particular, taking $\alpha$=0 as an example, we show the results in Table \ref{tab:IG}, and have the following observations:

\begin{table}[htbp]\small
    \renewcommand\arraystretch{1.3}
	\centering
	\caption{Impact of GCNSLIM optimization measures}
	\label{tab:IG}
	\begin{tabular}{c|cc|cc}
		\bottomrule[1.1pt]
		\multirow{2}[1]{*}{Model} & \multicolumn{2}{c|}{Wuhou} & \multicolumn{2}{c}{Ml-1M} \\
		\cline{2-5}    \multicolumn{1}{c|}{} & \multicolumn{1}{p{4.19em}}{Recall@10} & \multicolumn{1}{p{4.19em}|}{NDCG@10} & \multicolumn{1}{p{4.19em}}{Recall@10} & \multicolumn{1}{p{4.19em}}{NDCG@10} \\
		\hline
		GCNSLIM+0 & 0.5922 & 0.5888 & 0.1724 & 0.3642 \\
		GCNSLIM-LR & 0.5884 & 0.5866 & 0.1774 & 0.3728 \\
		GCNSLIM+0-LR & 0.5903 & 0.5867 & 0.1716 & 0.3628 \\
		GCNSLIM & \textbf{0.5940} & \textbf{0.5898} & \textbf{0.1778} & \textbf{0.3752} \\
		\toprule[1.1pt]
	\end{tabular}%
	\label{tab:addlabel}%
\end{table}%

(1) GCNSLIM consistently outperforms other variants, empirically showing that GCNSLIM can significantly benefit from the two proposed optimization measures. This implies that both the weakening of item self-similarity by removing layer 0 embeddings and the adaptation for strong centrality nodes by adding nonlinear activation can effectively improve the recommendation performance of GCNSLIM.\\
(2) On the Wuhou dataset, the performance degradation of GCNSLIM-LR compared with GCNSLIM is significantly greater than that of Ml-1M dataset. This implies that in the e-government service recommendation scenarios with a larger user/item volume gap, the centrality of item nodes that SLIM-like methods focus on is stronger, and nonlinear activation is more required.\\
(3) On the Ml-1M dataset, the performance degradation of GCNSLIM+0 compared to GCNSLIM is significantly greater than that of the Wuhou dataset. This implies that in recommendation scenarios where the number of users/items is more balanced and the number of interactions is more abundant, SLIM-like methods need to place greater emphasis on removing item self-similarity, so as to fully learn the collaborative signals between items.

\begin{figure}[h]
	\centering
	\subfloat[\label{fig:6a}]{
		\includegraphics[scale=0.45]{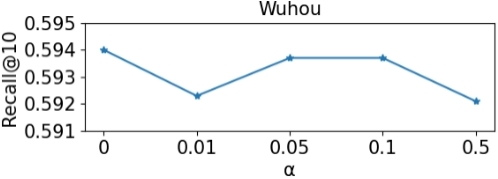}}
	\\
	\subfloat[\label{fig:6b}]{
		\includegraphics[scale=0.45]{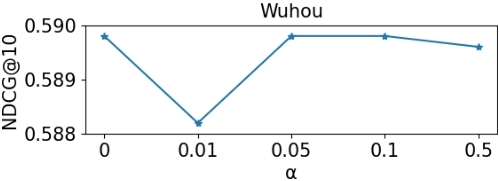}}
	\\
	\subfloat[\label{fig:6c}]{
		\includegraphics[scale=0.4]{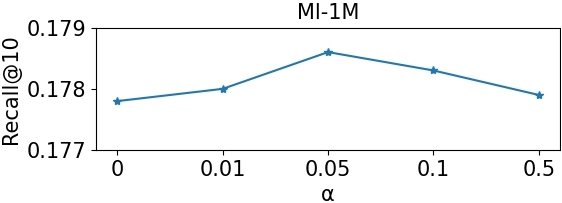}}
	\\
	\subfloat[\label{fig:6d}]{
		\includegraphics[scale=0.45]{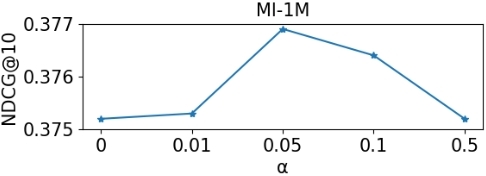}}
	\caption{GCNSLIM’s performance on different $\alpha$.}
	\label{fig6} 
\end{figure}
\subsection{Hyperparameter Experiments}
\subsubsection{The impact of joint optimization weight $\alpha$}
To analyze the impact of the joint optimization weight on GCNSLIM, we vary the joint optimization weight $\alpha$ in \{0, 0.01, 0.05, 0.1, 0.5\} and show the results in Figure \ref{fig6}. As shown in Figure \ref{fig6}, the recommendation performance of GCNSLIM on the Wuhou dataset reaches the best when $\alpha$=0, indicating that in the e-government service recommendation scenarios with a larger user/item volume gap, GCN with the pure SLIM framework can achieve the best recommendation performance. On the Ml-1M dataset, the recommendation performance of GCNSLIM first increases and then decreases with the increase of $\alpha$, and the performance reaches the best when $\alpha$=0.05. This shows that the joint optimization measure in scenarios with a more balanced user/item volume can effectively improve the recommendation performance of GCNSLIM, but too high joint optimization weight may lead to model overfitting and damage performance.

\begin{figure}[h]
	\centering
	\subfloat[\label{fig:7a}]{
		\includegraphics[scale=0.5]{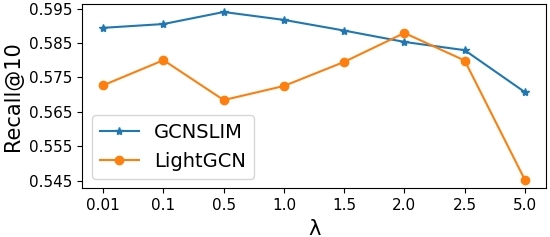}}
	\\
	\subfloat[\label{fig:7b}]{
		\includegraphics[scale=0.5]{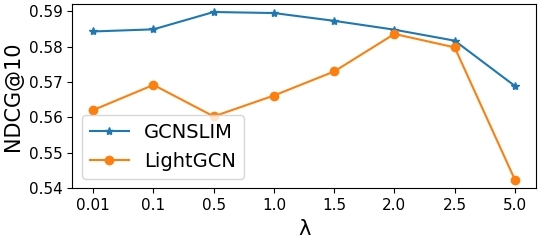}}
	\caption{Performance of GCNSLIM and LightGCN on different $\lambda$.}
	\label{fig7} 
\end{figure}

\subsubsection{The impact of L2 regularization rate $\lambda$}
To analyze the effect of L2 regularization rate $\lambda$ on GCNSLIM, taking $\alpha$=0 as an example, we change the L2 regularization rate $\lambda$ on the Wuhou dataset to \{0.01, 0.1, 0.5, 1.0, 1.5, 2.0, 2.5, 5.0\} and report the results in Figure 7. As shown in Figure \ref{fig7}, GCNSLIM is relatively insensitive to $\lambda$, and its recommendation performance trend is more stable than that of LightGCN. This shows that GCNSLIM is more immune to overfitting—because its core training parameter is only the item ID embeddings of layer 0, which is easier to train and regularize than LightGCN that pays equal attention to user ID embeddings and item ID embeddings. Overall, the optimal values of $\lambda$ for GCNSLIM and LightGCN are 0.5 and 2.0, respectively. When $\lambda$ is greater than 2.5, the performance of both models decreases rapidly. It shows that too strong regularization will negatively affect the normal training of models.
\par Overall, our comprehensive experimental results show that our GCNSLIM is an efficient and robust graph collaborative filtering recommendation algorithm, which can flexibly adapt to application scenarios such as e-government service recommendation to obtain higher recommendation accuracy and ideal operating performance.

\section{CONCLUSION AND FUTURE WORK}
In this work, we propose GCNSLIM for e-government service recommendation scenarios, which innovatively integrates light GCN on the SLIM framework, and effectively combines the advantages of both — SLIM’s efficient modeling ability for unbalanced user/item volume, and GCN’s ability to capture high-order collaborative signals. Additionally, we also propose two optimization measures — removing layer 0 embeddings and adding nonlinear activation, which further adapt to the characteristics of e-government service recommendation scenarios. Furthermore, we propose a joint optimization mode to flexibly balance the advantages of SLIM and MF to adapt to more diversified recommendation scenarios. Our extensive experiments on a real-world e-government service dataset and a commonly used public dataset demonstrate the rationality and effectiveness of combining GCN and SLIM.
\par We believe the insights from GCNSLIM are instructive for the future development of e-government service recommendation models. This work is an initial attempt to combine the advantages of both GCN and SLIM in collaborative filtering recommendation, and opens new research possibilities. In particular, in terms of GCN, we can learn a more uniform distribution of users and items to alleviate the popularity bias by designing a contrastive learning mode \cite{cai2023lightgcl} suitable for e-government service recommendation scenarios; We can integrate the attention mechanism \cite{song2019autoint,yang2023dgrec} into the message or layer aggregation functions to assign varying weights to different messages or layer embeddings based on their importance, which can improve the interpretability of recommendation results; In terms of SLIM, we can try to adopt multinomial likelihood loss function \cite{liang2018variational} or contrastive loss function (e.g., cosine contrastive loss \cite{mao2021simplex}) instead of simple squared loss to achieve higher ranking accuracy. Furthermore, with the continuous development of e-government services, the number of active users with rich interaction history will gradually increase, and the contextual data of users and items will also be gradually accumulated. At that time, in addition to GCN and other advanced general recommendation technologies can be introduced into the field of e-government service recommendation, advanced sequence recommendation \cite{hou2022core,fan2021lighter,pan2020star} and context-aware recommendation \cite{song2019autoint,wang2021dcn} technologies (such as Bert, SAN, etc.) will also have great application space in the field of e-government service recommendation.


\bibliographystyle{ACM-Reference-Format}
\bibliography{sigconf}

\appendix

\end{document}